\documentclass[a4paper, 10pt]{article}
\usepackage{graphicx}
\newcommand{\beq}{\begin{equation}}
\newcommand{\eeq}{\end{equation}}
\newcommand{\beqa}{\begin{eqnarray}}
\newcommand{\eeqa}{\end{eqnarray}}

\def\half{\frac{1}{2}}

\def\ket#1{|#1\rangle}
\def\bra#1{\langle\, #1\,|}

\def\opone{\leavevmode\hbox{\small1\normalsize\kern-.33em1}}

\title{A possible definition of a {\it Realistic} Physics Theory}

\author{Nicolas Gisin \\
\it \small   Group of Applied Physics, University of Geneva, 1211 Geneva 4,    Switzerland}

\date{\small \today}

\begin{document}
\maketitle

\begin{abstract}
A definition of a {\it Realistic} Physics Theory is proposed based on the idea that, at all time, the set of physical properties possessed (at that time) by a system should unequivocally determine the probabilities of outcomes of all possible measurements.
\end{abstract}

\section{Introduction}\label{intro}
Some of the physics literature of the last two decades abundantly mentions {\it realistic} theories, especially when discussing quantum physics, nonlocality and Bell inequalities (see e.g. \cite{Zeilinger2007, Zeilinger2013, Weinfurter2012, Kwiat2013}), or epistemic versus ontic interpretations of the quantum state \cite{Spekkens10,Pusey2012}. On the one side, this is highly interesting as most physicists have some intuitive understanding of what "reality" is. However, on the other side, it is very disappointing, because the vast majority of this literature doesn't properly define {\it realistic} theories and thus obscures what could be an interesting debate \cite{GisinScience09}.

The aim of this note is to propose a plausible definition of a {\it realistic theory}, section \ref{def}. It is not entirely new, but seems to have been forgotten. A good definition should be testable, however, the definition I propose is not testable in one single experiment. Nonetheless it defines whether a given physics theory is realistic or not. Hence, testing our best realistic physics theories amounts to testing whether we can maintain a realistic physical description of nature. Section \ref{NL} discusses nonlocality in the context of our definition.

\section{Definition of {\it Realistic} Physics Theories}\label{def}
In the physics literature one finds essentially two characterizations of {\it realistic} theories:

\begin{enumerate}
\item All measurement outcomes are determined by the state of the physical system.  In other words, at any time all physical quantities have their value somehow written in the physical system (these may change as time passes).

\item All measurement outcome probabilities are determined by the state of the physical system.  In other words, at any time all physical quantities have the probabilities of all their possible values written in the physical system (again, these may change as time passes).

\end{enumerate}

Note that both characterizations incorporate the idea that measurement outcomes (or their probabilities) are defined prior and independently of the observer, as one would expect for a realistic theory. Nevertheless, both characterizations are unsatisfactory. The first one identifies {\it realistic} with {\it deterministic}. Indeed, if measurement outcomes are {\it written in the system}, then, either randomness can only be due to our ignorance of the precise state of the system, i.e. randomness is not intrinsic but merely epistemic, or randomness is merely due to non-perfect measurements, i.e. randomness is technical as in classical physics. The second characterization is not better, as one can hardly see how a theory could not satisfy it, i.e. how the state of a physical system could not determine the probabilities of all measurement outcomes. Indeed, such "states" would make no prediction at all, they would not even predict the frequencies of measurement outcomes\footnote{This is true whatever interpretation of probabilities one uses; indeed, whatever probabilities are, at the end of the day, probabilities are tested by comparing them with frequencies in long series of measurements.}. In brief, no physicist would call such "states", states of physical systems.

Consequently, the first characterization is too strong, as it imposes determinism, while the second characterization is too weak, as all scientific theories have to satisfy it. However, intuitively, both characterizations contain some truth. The state of physical system must determine the probabilities of all possible measurement outcomes, as suggested by the second characterization. And at least some physical properties\footnote{We distinguish physical quantities, like e.g. energy, and physical properties, like e.g. an energy eigenprojector. When testing a property the answer is binary, yes/no. A physical quantity corresponds to an entire set of compatible properties. In classical physics properties are represented by subsets of the phase-space, while in quantum theory properties are represented by projectors.} must be written in the physical system, as suggested by the first characterization (if not how would one even recognize the system?). This leads to the following definition:\\

{\bf Definition 1:} A theory is {\it realistic} if and only if, according to the mathematical structure of this theory, the collection of all physical quantities written in the system unambiguously determines the probabilities of all possible measurement outcomes.\\

In other words, the theory is such that, at all times, the collection of all properties possessed\footnote{In this note I use equivalently the following terminologies: possessed property, actual property, property written in the system.} by the system fully determines all relevant probabilities: the collection of physical quantities without uncertainty (or, more precisely, without indeterminacy) determines the probabilities of all undetermined physical quantities. Notice that a possessed property is nothing but what EPR called an element of reality \cite{EPR35}.

This definition captures both the idea that, at any time, some properties (physical quantities) are written in the system, as suggested by characterization number 1, and that what is written in the system determines all the relevant probabilities, as suggested by characterization number 2. When time passes, some properties possessed by the system pass away, while the system acquires new properties, i.e. the state of the system evolves.

Let us present our definition more rigorously (see also \cite{GisinPropensity91}). Our definition assumes that physics theories have some concept of {\it physical properties}, here denoted $\pi$, and of {\it probability measures}, $\omega: \pi \rightarrow \omega(\pi)\in [0..1]$. Furthermore, it is the theory that says which sets of properties $\{\pi_j\}$ can be measured jointly in a single measurement (whose outcomes corresponding to the - thus compatible - $\pi_j$). Probability measures must satisfy the following two natural conditions:
\begin{enumerate}
\item If, according to the theory, a given measurement has several possibly mutually exclusive outcomes $\pi_j$, then, grouping several outcomes together\footnote{The outcome $\Pi$ corresponds to ``one of the $\pi_j$ happened''.}, denoted $\Pi\equiv\{\pi_j\}$, one has $\omega(\Pi)=\sum_j \omega(\pi_j)$.
\item If, for any measurement, one groups together all possible outcomes, then $\omega(\opone)=1$, where $\opone$ denotes the property corresponding to grouping together all outcomes.
\end{enumerate}
A physics theory must also have some concept of {\it states}, here denoted $\phi$. Clearly, every state $\phi$ should determine a unique probability distribution $\omega_\phi$. However, one should not identify states with probability measures. Indeed, if only the probabilities $q_j$ of possible states $\phi_j$'s are known, then the probability measure reads $\sum_j q_j \omega_{\phi_j}$. Hence, not all probability measures correspond to a (pure) state. 

If $\omega_\phi(\pi)=1$, then, when the system is in state $\phi$, the outcome $\pi$ is written in the physical system (here, a unit probability is identified with certainty). 

We can now present our definition mathematically. A theory is realistic iff, according to the mathematical structure\footnote{More precisely, the mathematical structure of the set of all properties, according to this theory.} of this theory, the following condition holds for all (pure) states $\phi$ and all probability measures $\omega$:\\

If $\{\pi|\omega(\pi)=1\}=\{\pi|\omega_\phi(\pi)=1\}$, then $\omega=\omega_\phi$.\\ 

Hence, in realistic theories the state of a physical system can be identified with the collection of possessed properties. Indeed, this collection determines unequivocally all relevant probabilities.

This definition is in accordance with the idea that the state of a physical system is the collection of all "actual properties", i.e. of all physical quantities such that if one measures them, the outcome is certain or predetermined \cite{Piron1976, Piron1Lesson}. Accordingly, for any deterministic theory the state is merely a summary of all possible measurement results, as in classical mechanics and in some interpretations of quantum mechanics like Everett \cite{Everett} and Bohm\footnote{Notice that in Bohm's theory two apparata corresponding to the same self-adjoint operator and identical initial state-vector, but with different hidden positions of the particles making up the apparata may lead to different measurement outcomes (this is true even for position measurements \cite{Werner09}). Accordingly, in Bohm's theory there are many more properties than in quantum theory \cite{NaiveBohm}. Consequently, the set of possessed properties in Bohm's theory defines not only the state-vector of the system, but also its hidden positions.} \cite{BohmPilotWave}. For non-deterministic theories, the situation is more interesting \cite{GisinPropensity91}. For example, according to quantum theory, at any time $t$, there are physical quantities, represented in this theory by self-adjoint operators $A_j$, with well determined values $a_j$: $A_j\Psi_t=a_j\Psi_t$, where $\Psi_t$ represents the quantum state at time $t$. Properties are represented by projectors, e.g. the eigenprojectors $\pi_j^a$ of the operators $A_j$ with spectral decompositions $A_j=\sum_{a_j} a_j\cdot\pi_j^a$.

Noteworthy, for any (normalized) pure state $\Psi_t$, the collection of all properties possessed by the system, $\{\pi^a_j|A_j\Psi_t=a_j\Psi_t\}$, determines a unique probability distribution $\omega_{\Psi_t}$ such that $\omega_{\Psi_t}(\pi^a_j)=1$ for all outcomes $\pi^a_j$ corresponding to a possessed property. This is nothing but Gleason's theorem (valid in all Hilbert spaces of dimension $\ge3$ \cite{PeresBook}). As is well-known, this probability distribution reads $\omega_{\Psi_t}(\pi_j^a)=\bra{\Psi_t}\pi_j^a\ket{\Psi_t}$.

According to Definition 1 both classical and quantum theories are realistic. Note that, in both theories, mixed states do not satisfy the proposed definition. This should be no surprise, as mixed states describe situations involving some ignorance, either about the precise state preparation or about a part of the global system which has been ignored (traced out, as physicists say).

The case of 2-dimensional quantum systems, so-called qubits, is interesting. Pure states can be labeled by vectors pointing on the unit sphere in ordinary 3-dimensional space, $\phi=\ket{\vec m}$, and all nontrivial properties, (i.e. not equal to the identity $\opone$) can also be labeled by vectors on the unit sphere $\pi_{\vec p}=\ket{\vec p}\bra{\vec p}$. Compatible (nontrivial) properties correspond to antipodal points on the sphere. According to quantum theory, every state $\ket{\vec m}$ is associated to the following probability measure: $\omega_{\vec m}(\pi_{\vec p})=\bra{\vec m}\pi_{\vec p}\ket{\vec m}=\frac{1+\vec m\cdot\vec p}{2}$. Accordingly, for every state $\ket{\vec m}$ there is a unique (nontrivial) actual property $\pi_{\vec m}$ that corresponds to the same point on the sphere. However, since Gleason's theorem doesn't hold in dimension two, there are different probability measures with precisely the same set of actual properties. For example, the probability measure\footnote{Note that $\omega(\pi_{\vec p})+\omega(\pi_{-\vec p})=\omega(\opone)=1$, as it should.} $\omega(\pi_{\vec p})=\frac{1+(\vec m\cdot\vec p)^3}{2}$ takes the value one on precisely the same set of properties as $\omega_{\vec m}$, but clearly differs from $\omega_{\vec m}$. Such an example is impossible for qutrits and all quantum systems described by Hilbert spaces of dimensions larger or equal to 3.

Consequently, quantum theory limited to one qubit is an example of a non-realistic theory. But, clearly, qubits don't exist per se, in some disembodied form, but only as subsystems of larger - actually infinite dimensional - systems; hence this is a nice example, but not a compelling one against fundamental theories being realistic.

Let me conclude this section by admitting that I would be greatly surprised, even shocked, if any future fundamental theory in physics would be non-realistic according to the proposed definition\footnote{For me, realism is merely the existence of a world outside me and independent of me. A priori, this world could be described as well by classical physics as by quantum theory or by any future theory. Indeed, the existence of an outside world doesn't impose any restriction on how this world is (the main and only point is that it exists). This outside world could be deterministic or random, local or nonlocal, entirely within space-time or not, fully understandable by our intelligence or not. Hence, physics will never be able to deny its existence. However, let me confess an act of faith: I believe that by studying physics I open a small skylight on this external world. What I can see is only a tiny part of it, but already this tiny part is absolutely fascinating: enlarging this small window is the great Enterprise of Science. And let me add that if one assumes realism, then it is all the most natural to look for realistic theories and to be satisfied only by realistic theories.}. But logically this is certainly possible.

\section{Nonlocality}\label{NL}
In the context of quantum nonlocality, some physicists conclude that no theory compatible with the violation of Bell inequalities can be realistic. As long as the word "realistic" is not defined, one can't argue against (nor in favor) of such a statement \cite{nonrealism}. According to  Definition 1, such a statement is clearly wrong, as quantum theory predicts the violation of some Bell inequalities and is realistic. However, in order for a realistic theory to predict the violation of some Bell inequalities, the theory must incorporate some form of nonlocality. Definition 1 of a {\it realistic theory} rests on the concept of {\it possessed properties}, or equivalently of {\it physical quantities without indeterminacy}. Thus, if the system is delocalized, like a delocalized photon in two (or more) optical modes or like two (or more) entangled particles located at a distance from each other, then some of the physical quantities without indeterminacy are also delocalized. Note that in order to measure such delocalized physical quantities the measurement apparatus itself must be delocalized or must consist of several parts. It is certainly highly counter-intuitive that some properties written (at some point in time) in a physical system can be delocalized, but that doesn't make them less real. For example, if two systems have their spin $\half$ entangled as in the singlet state, then their global spin is zero although this global spin is delocalized; one way to measure it uses an apparatus with two pieces, one near each particle, each measuring the spin $\half$ along the same direction; the pre-determined outcome is that the two local results are opposite. In this example, the possessed property {\it global spin zero} is written in the 2-particle system, hence it is written out there in space, but in a nonlocal fashion.

\section{Conclusion} \label{concl}
I proposed a possibly definition of realistic theories which avoids the mere reduction to determinism and avoids the risk of tautology. This definition is closely related to the concept of state as the collection of all actual properties, or equivalently as the collection of all elements of reality. According to it, both classical and quantum theories are realistic. Consequently, one can't avoid nonlocality by merely rejecting realistic theories. Still, non-realistic theories exist mathematically and one may wonder whether any future fundamental physics theory may be non-realistic.

Other definitions of realistic physics theories are possible \cite{BubClifton}. It is an interesting research program to compare them and to investigate the status of Bell-nonlocality in each of them.




\section*{Acknowledgment}
It is a pleasure to thank Tomer Barnea, Jeff Bub, Michael Esfeld, Yeong-Cherng Liang, Gilles Puetz, Terry Rudolf and Valerio Scarani  for helpful comments.

\end{document}